\documentclass[11pt,aps,floats,showpacs,amssymb,tightenlines,nofootinbib]{revtex4}
\usepackage{color}
\usepackage{graphicx}
\usepackage{amsmath}
\usepackage{amsfonts}
\usepackage{amscd}
\usepackage{epsfig}
\usepackage{amssymb}
\usepackage{tabularx}

\begin{document}

\title[]{Vacuum thin shells in Einstein-Gauss-Bonnet brane-world cosmology}

\author{Marcos A. Ramirez$^{1,2}$}

\affiliation{$^1$ Instituto de F\'{i}sica Enrique Gaviola - CONICET, FaMAF - Universidad Nacional de C\'ordoba, (5000) C\'ordoba, Argentina}
\affiliation{$^2$ Instituto de Investigaciones en Energ\'{i}a no Convencional - CONICET, Facultad de Ciencias Exactas - Universidad Nacional de Salta, (4400) Salta, Argentina}

\begin{abstract}
In this paper we construct new solutions of the Einstein-Gauss-Bonnet field equations in an isotropic Shiromizu-Maeda-Sasaki brane-world setting which represent a couple of $Z_2$-symmetric vacuum thin shells splitting from the central brane, and explore the main properties of the dynamics of the system. The matching of the separating vacuum shells with the brane-world is as smooth as possible and all matter fields are restricted to the brane. We prove the existence of these solutions, derive the criteria for their existence, analyse some fundamental aspects or their evolution and demonstrate the possibility of constructing cosmological examples that exhibit this feature at early times.
We also comment on the possible implications for cosmology and the relation of this system with the thermodynamic instability of highly symmetric vacuum solutions of Lovelock theory.

\end{abstract}

\maketitle
\section{Introduction}

Lovelock's theory of gravity is arguably the most natural higher-dimensional generalisation of general relativity \cite{lovelock}. For a 4-dimensional spacetime, Lovelock gravity is precisely general relativity plus an eventual cosmological constant. In the case of 5 or 6-dimensional spacetimes, this theory constitutes the so-called Einstein-Gauss-Bonnet (EGB) gravity, whose action differs from that of general relativity in the addition of a term of second order in the curvature, namely
\begin{equation}
I=\frac{1}{2\kappa^2}\int d^D x \sqrt{-g}(R-2\Lambda+\alpha L_{GB}) + I_m \;\; , \;\; L_{GB}=R^2-4R_{\mu\nu} R^{\mu\nu}+R_{\alpha\beta\mu\nu}R^{\alpha\beta\mu\nu} ,
\end{equation}
where $I_m$ stands for the action of the extant matter-energy fields. EGB gravity also appears as a classical limit of certain string theories \cite{zwiebach}. Taking into account this stringy motivation, here we will consider $\alpha>0$.  

A relatively simple way to obtain non-vacuum solutions of a given set of field equations, that are of physical significance, is through the introduction of thin shells. These are hypersurfaces which represent a concentrated source for the field: matter-energy concentrated on a codimension one submanifold. Mathematically, these objects are characterised by junction conditions, which, in the case of gravity, are equations that relate the discontinuity of the extrinsic curvature of the shell with the intrinsic stress-energy tensor defined on the submanifold. In the case of Einstein-Gauss-Bonnet theory, provided the only source for the gravitational field is the thin shell, the junction conditions are \cite{junctionEGB}
\begin{equation}
\label{junction}
[Q^a_b]_{\pm}=-\kappa^2 S^a_b \;\;\; , \;\;\; Q^a_b:=K^a_{b}-\delta^a_bK+2\alpha(3J^a_b-\delta^a_bJ-2P^a_{cbd}K^{cd}),
\end{equation}
where the brackets represent the difference at both sides of the shell of the quantity they enclose, 
\begin{eqnarray}
J_{ab}&:= &\frac{1}{3}(2KK_{ad}K^d_b + K_{df}K^{df}K_{ab} - 2K_{ad}K^{df}K_{fb}- K^2K_{ab}), \\
P_{adbf}&:= &R_{adbf} + 2h_{a[f}R_{b]d} + 2h_{d[b}R_{f]a} + Rh_{a[b}h_{f]d},
\end{eqnarray}
and $S_{ab}$ is the intrinsic stress-energy tensor on the shell (all these tensors are defined within the submanifold). It is important to notice the possibility of having non-trivial solutions ($[K^a_b]\neq0$) even in the case $S_{ab}=0$: these are the so-called {\it vacuum thin shells} \cite{gravanis} \cite{garraffo}, which are vacuum solutions of low regularity ($C^0$ at the shell) nonexistent in general relativity.

A relevant and relatively recent application of thin shells is {\it braneworld cosmology} (see \cite{maartens} for a review). In this setting the observable universe is a $4$-dimensional thin shell (braneworld), in which all the standard model fields live, embedded in a higher-dimensional spacetime (bulk) that is usually asymptotically $AdS_5$. It has attracted considerable interest because it is inspired by results in $M$-theory, offers an alternative (to compactification) explanation regarding the invisibility of extra dimensions and the hierarchy problem \cite{randallsundrum}, and it can be used to construct models that reproduce standard cosmology including dark energy and inflation. Although the classical limit for the $5$-dimensional gravity theory is usually regarded to be general relativity, EGB gravity is more general and, as mentioned, it is also a classical (low-energy) limit of certain string theories. In this way it has been applied to the context of braneworld cosmology \cite{charmousisdufaux} and different aspects of the dynamics have been analysed (see \cite{EGBbraneworld,maeda} and references therein).



On the other hand, a new type of stability analysis for thin shells has been recently developed \cite{ramirez}. It consists on an infinitesimal separation into two parts of the constituent matter-energy fields, configuring in this way two different shells with an intermediate bulk, where the resulting spacetime can be determined by continuity of the normal vector of the shell (both splitting shells have the same initial normal vector). It can be understood as a way to determine how well are these constituents gravitationally confined within a single shell. In this work we propose to use this analysis in an EGB braneworld context but with one important difference: we are not going to separate constituents of the brane, we will consider vacuum thin shells {\it emanating} from a given braneworld solution, which radically changes the bulk while leaving the brane with the same matter-energy content. As we will show, this analysis turns out to be non-trivial and will demonstrate the existence of a new class of solutions in the context of EGB gravity not previously analysed in the literature.

We begin with a derivation of the equations of motion of the different shells involved in this construction: the central brane-world in Section \ref{brane} and the separating vacuum thin shells in Section \ref{vacuum}. Then, in Section \ref{splitting}, we obtain criteria that determine the existence of this kind of solution, and prove that the criteria are satisfied for a range of parameters. The possible final outcomes of the evolution are considered in Section \ref{outcome}. Finally, in Section \ref{example} we give an example that tends to our universe in the limit of large scale factor, and in section \ref{conclusions} we summarise our results, propose possible interpretations, discuss the physical relevance of the solution we found and compare with other results in the field of Lovelock gravity.

\section{Isotropic thin shell with $Z_2$-symmetry}
\label{brane}

Let us consider a $4$-dimensional timelike thin shell made of a perfect fluid embedded in a $Z_2$-symmetric $5$-dimensional vacuum bulk spacetime that is placed at the symmetry centre. As usual in braneworld contexts, there is positive brane tension $\sigma>0$ on the thin shell. We also impose that the spacetime is foliated by $3$-dimensional constant curvature spacelike submanifolds\footnote{This scenario is usually called SMS, or Shiromizu-Maeda-Sasaki braneworld \cite{SMS}.}, so the metric of any of the identical bulk regions is given by \cite{zegers}
\begin{equation}
\label{metric}
ds^2=-f(r)dt^2+f^{-1}(r)dr^2+r^2d\Sigma_k^2 \;\;\;\; , \;\;\;\; f(r)=k+\frac{r^2}{4\alpha}\left(1+\xi\sqrt{1+\frac{4\alpha\Lambda}{3}+\frac{\mu\alpha}{r^4}}\right)
\end{equation}
where $d\Sigma_k^2$ stands for the metric of the corresponding constant curvature manifold ($k=-1,0,1$), $\xi=\pm 1$ (the ``minus" branch is the so-called general-relativistic (GR) branch, while the ``plus" one is the stringy branch), and $\mu$ is the mass parameter. In order to have an asymptotic limit for large $r$, we will impose $1+4/3\alpha\Lambda>0$. We then define $\beta=\sqrt{1+4/3\alpha\Lambda}$. One can see that in this limit, for $\Lambda\neq0$, the metric tends to de Sitter or Anti-de Sitter depending on the value of the ``effective cosmological constant''
\begin{equation}
\Lambda_{eff} =-\frac{3}{2\alpha}\left(1+\xi\beta \right) .
\end{equation}
From this equation it is deduced that the sign of $\Lambda_{eff}$ for the GR branch is the same as the sign of $\Lambda$ (as $\alpha>0$), while the stringy branch is always asymptotically $AdS$. In particular, if $\Lambda=0$ then the GR branch is asymptotically flat, while the other is asymptotically $AdS$. From (\ref{metric}) it is also deduced that if $\mu=0$ then the solution is maximally symmetric. Then $1+4/3\alpha\Lambda>0$ is also a necessary condition to have maximally symmetric solutions.

On the other hand, the intrinsic metric of the shell is given by
\begin{equation}
ds^2_{\cal{S}}=-d\tau^2+a(\tau)^2d\Sigma_{k}^2 ,
\end{equation}
where $\tau$ is the proper time of the shell. Because of the symmetries, the intrinsic stress-energy tensor can be written as $S_i^j=diag[-\rho,p,..,p]$. We impose that the matter content of the brane satisfies the dominant energy condition. Applying the junction conditions (\ref{junction}), we obtain two independent equations: one that relates the energy density within the shell $\rho$ with bulk parameters, the scale factor of the shell and its first derivative (the $\tau \tau$ component); while the other relates the pressure within the shell $p$ with bulk parameters, the scale factor and its first two derivatives (any of the diagonal spacelike components). In a non-static situation ($\dot{a}\neq 0$), the second equation is a consequence of the first one and the conservation of the source ($S^j_{i;j}=0$), so we will focus on the $\tau \tau$ component of the junction conditions and the continuity equation. Explicitly, at any given side of the shell we have
\begin{equation}
\label{Q}
Q^{\tau}_{\tau}=-sign\left(\left.\frac{\partial r}{\partial\eta}\right|_{\eta=0^+}\right)\frac{3}{a^3}\sqrt{\dot{a}^2+f(a)}\left(a^2+4\alpha\left(k+\frac{2}{3}\dot{a}^2-\frac{1}{3}f(a)\right)\right),
\end{equation}
where $\eta$ is the gaussian normal coordinate of the shell, and we are evaluating this quantity at the $\eta>0$ side\footnote{Because of the $Z_2$ symmetry, at the $\eta<0$ side the expression for $Q^{\tau}_{\tau}$ is exactly the same but with the opposite sign, stemming from $\left.(\partial r/\partial \eta)\right|_{\eta=0^-}$.}.
Taking squares, the $\tau \tau$ component of (\ref{junction}) implies \cite{maeda}
\begin{equation}
\label{rhosq}
\frac{\kappa^2}{36}(\rho+\sigma)^2=\left(\frac{f(a)}{a^2}+H^2\right)\left[1+\frac{4\alpha}{3}\left(\frac{3k-f(a)}{a^2}+2H^2\right)\right]^2,
\end{equation}
where $H=\dot{a}/a$. This equation is equivalent to that component of (\ref{junction}) only if the orientation for the $r$ coordinate of the bulk is in agreement with (\ref{orientation}), as explained in Appendix \ref{derivation}, which in this case implies that the bulk should be interior.
From (\ref{rhosq}), an equation of motion for the shell can be derived, provided $\alpha>0$, (see \cite{maeda} and Appendix \ref{derivation})
\begin{equation}
\label{branemotion}
H^2+V_{\mu,\xi}(a)=0,
\end{equation}
where
\begin{equation}
V_{\mu,\xi}(a)=-\frac{1}{8\alpha}\left[B_{\xi}(P(a)^2,A_{\mu}(a)^{3/2})+\frac{A_{\mu}(a)}{B_{\xi}(P(a)^2,A_{\mu}(a)^{3/2})}-2-\frac{8k\alpha}{a^2} \right],  
\end{equation}
\begin{equation}
B_{\xi}(P^2,A_{\mu}^{3/2})=-\xi A_{\mu}^{3/2}+256\alpha^3P^2+16\sqrt{2\alpha^3P^2(128\alpha^3P^2-\xi A_{\mu}^{3/2})} ,
\end{equation}
and the functions $P(a)$ and $A_{\mu}(a)$ are defined by
\begin{equation}
\label{AP}
P(a)=\frac{\kappa^2}{16\alpha}\left(
\rho+\sigma \right) \;\;\;\; , \;\;\;\; A_{\mu}(a)=\beta^2+\frac{\alpha\mu}{a^4}.
\end{equation}
In order for $B_{\xi}$ to be well defined it is necessary to have $A_{\mu}>0$ and $128\alpha P^2-\xi A_{\mu}^{3/2}>0$ (provided $\rho+\sigma \neq 0$), so we will assume these from now on, and consequently $B_{\xi}>0$ must hold. In this way, equation (\ref{branemotion}) together with a function $\rho(a)$ characterising the matter-energy content of the brane determines $a(\tau)$ for a given initial data $(a(\tau_0), sign(\dot{a}(\tau_0))$. Alternatively, the function $\rho(a)$ can be obtained by solving the continuity equation 
\begin{equation}
\label{continuity}
\frac{d\rho}{da}+\frac{3(\rho + p)}{a}=0,
\end{equation}
provided a barotropic equation of state $e(\rho,p)=0$ is given.   


\section{Vacuum thin shell}
\label{vacuum}

One can notice the possibility that equation (\ref{junction}) may have non-trivial solutions if the right hand side is zero. It is known that this is indeed the case, in EGB gravity there exist vacuum thin shells \cite{gravanis}. These shells can be understood as an interface between two different vacuum solutions, and as a weak solution of the vacuum field equations\footnote{To my knowledge, this interpretation lacks a formal proof.}. The properties of this kind of shells in this setting have been thoroughly analysed in \cite{garraffo}, and here we are only summarising the ones important to our purpose. In a spacetime with the symmetries we imposed, an equation of motion for the vacuum shell can be obtained, namely
\begin{equation}
\dot{a}^2+V_{vac}(a)=0 ,
\end{equation}
where the potential is given by
\begin{equation}
\label{vaceq}
V_{vac}(a)=k + \frac{a^2}{4\alpha} -\frac{a^2}{4\alpha}\left(\frac{3(\xi_+ A^{1/2}_{\mu_+} +\xi_- A^{1/2}_{\mu_-})^2+(\xi_+ A^{1/2}_{\mu_+} -\xi_- A^{1/2}_{\mu_-})^2}{12(\xi_+ A^{1/2}_{\mu_+} +\xi_- A^{1/2}_{\mu_-})}\right), 
\end{equation}
the subindexes $\pm$ denote the different vacuum solutions being glued at each side of the shell, and the $A_{\mu}$ functions are those defined by (\ref{AP}). In our case, we will only consider the gluing of solutions of the same action, which means that $\Lambda$ and $\alpha$ are the same at both sides. What can change from one side to the other are the mass coefficients $\mu_{\pm}$, the label of the branches $\xi_{\pm}$, and the orientation of the $r$ coordinate with respect to the shell (which means that the bulk regions can be either interior or exterior). 

As shown in \cite{garraffo}, the only way to possibly glue two GR-branches ($\xi_+=\xi_-=-1$) is by imposing that the construction has the wormhole orientation (which means that both solutions being glued should be exterior) and $\alpha<0$. Also, if the shell glues an interior solution with an exterior one, then they must correspond to different branches ($\xi_+\neq\xi_-$), that is, it must be a ``false vacuum bubble''. Also, for this configuration, the mass coefficients can not be equal ($\mu_+\neq\mu_-$).

Taking a first derivative of (\ref{vaceq}) one can show 
\begin{equation}
\label{accelvac}
\ddot{a}=-\frac{a}{4\alpha}\left(1-\frac{\beta^2}{\xi_+ A^{1/2}_{\mu_+} +\xi_- A^{1/2}_{\mu_-}}\right) .
\end{equation}
This expression will be useful in the next Section.

\section{Splitting construction and stability conditions}
\label{splitting}
\begin{figure}
\centerline{\includegraphics[width=.2\textwidth]{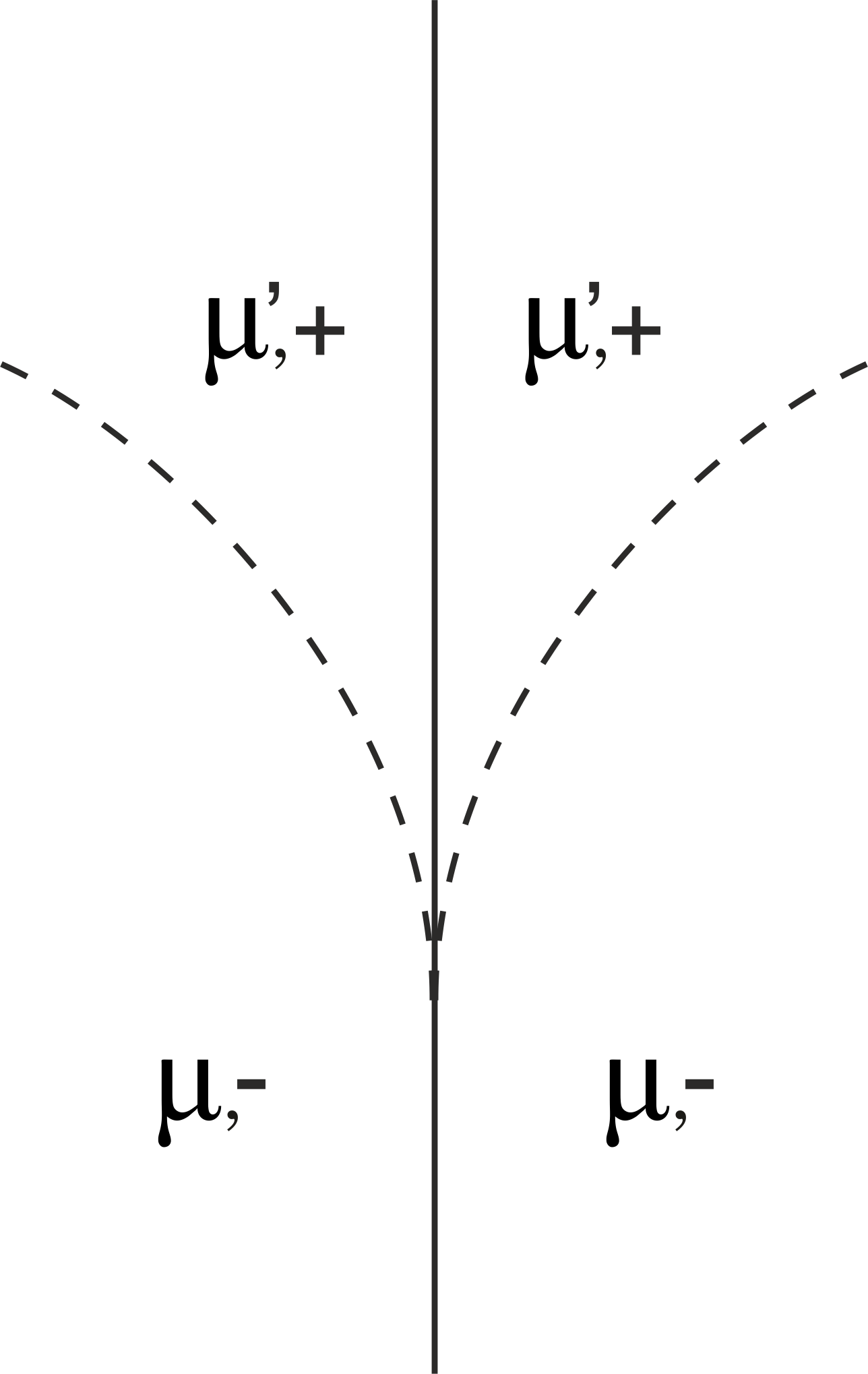}}
\caption{Illustration of the splitting solutions whose existence we are proving in this paper}
\label{vacuumsplitting}
\end{figure}

This is the novel part of the paper. Inspired by the possibility of constructing well-defined solutions of Einstein equations that represent splitting thin shells \cite{ramirez}, we explore the plausibility of the existence of a SMS brane-world solution with $Z_2$-symmetry from which a couple of vacuum thin shells emanate at a given point of the evolution, as illustrated by figure \ref{vacuumsplitting}. 

As the figure suggests, we consider an initial configuration in which the bulk space-time is originally of the GR type. Because of this fact and the assumption $\alpha>0$, the vacuum thin shells must be interfaces between a GR branch and a stringy branch. 
If we demand that the central brane-world must be an embedded submanifold everywhere, including the separation point, we then should impose that the normal vector of the brane is continuous (unique) at the separation point (which we characterise as $a_s$, the scale factor at that moment). 
This is the same as imposing continuity of $\dot{a}$ for the brane at this point, which can be written as $V_{\mu,-1}(a_s)=V_{\mu',1}(a_s)$. It turns out that this continuity condition implies that the normal vectors of the vacuum thin shells at the separation moment also coincide with that of the brane, so $V_{\mu',1}(a_s)=V_{vac}(a_s)$, where in (\ref{vaceq}) we would have $\xi_-=-1,\xi_+=1,\mu_-=\mu,\mu_+=\mu'$. This can be seen from the junction conditions (\ref{junction}) as $Q^{\tau}_{\tau}$, expressed in (\ref{Q}), is a function of $a$, $\dot{a}$ and the parameters that characterise the bulk $(\mu,\xi)$ at the side being analysed.
For the vacuum thin shell we have $Q^{\tau}_{\tau}(a_s,\dot{a}_{v,s},\mu,\xi=-1)=Q^{\tau}_{\tau}(a_s,\dot{a}_{v,s},\mu',\xi=1)$, where $\dot{a}_{v,s}$ is the derivative of the scale factor of the vacuum thin shell with respect to its proper time at $a_s$; while for the brane before and after the splitting we would have $2Q^{\tau}_{\tau}(a_s,\dot{a}_{b,s},\mu,\xi=-1)=2Q^{\tau}_{\tau}(a_s,\dot{a}_{b,s},\mu',\xi=1)=\kappa^2\rho(a_s)$, where $\dot{a}_{b,s}$ is the derivative of the scale factor of the brane with respect to its proper time also at $a_s$. Given the structure of $Q^{\tau}_{\tau}$, both equations can only hold if $\dot{a}_{v,s}=\dot{a}_{b,s}$, so we will call this common value $\dot{a}_s$.      

Furthemore, one can notice that $V_{\mu,-1}(a_s)=V_{\mu',1}(a_s)$ is equivalent to 
\begin{equation}
\label{continuity1}
B_{-1}(P^2(a_s),A_{\mu}^{3/2}(a_s))+\frac{A_{\mu}(a_s)}{B_{-1}(P^2(a_s),A_{\mu}^{3/2}(a_s))}=B_{1}(P^2(a_s),A_{\mu'}^{3/2}(a_s))+\frac{A_{\mu'}(a_s)}{B_{1}(P^2(a_s),A_{\mu'}^{3/2}(a_s))},
\end{equation}
which is an expression that allows one to set $\mu'$ as a function of $(\mu,a_s)$. Calculations are simpler if we define the functions
\begin{eqnarray}
x(a)=\frac{A^{3/2}_{\mu}(a)}{256\alpha^3P(a)^2} \;\; &,& \;\; y(a)=\frac{A^{3/2}_{\mu'}(a)}{256\alpha^3P(a)^2}  \\
\omega(x)=(1+x+\sqrt{1+2x})^{1/3} \;\;&,&\;\; \nu(y)=(1-y+\sqrt{1-2y})^{1/3}  \\
 g(x)=\omega(x)+\frac{x^{2/3}}{\omega(x)} \;\;&,&\;\; h(y)=\nu(y)+\frac{y^{2/3}}{\nu(y)} ,
\end{eqnarray}  
where $x(a)$ and $y(a)$ are non-negative. We also define $x_s=x(a_s)$, $y_s=y(a_s)$. Then, (\ref{continuity1}) can be written as
\begin{equation}
\label{continuity2}
g(x_s)=h(y_s).
\end{equation}
The function $\nu(y)$ is real only in the range $0\leq y \leq 1/2$. Nevertheless, $h(y)$ is real for any non-negative value of $y$, as it can be written as
\begin{equation}
\label{hextended}
h(y)=2y^{1/3} \cos(\alpha(y)/3)
\end{equation}
where $\alpha(y)=atan2(\sqrt{2y-1},1-y)$ is three times the argument of $\nu(y)$ when $y>1/2$ and is a monotonically increasing function of $y$ which tends to $\pi$ when $y\to\infty$. Furthermore, this expression also holds for $y\leq 1/2$ as this is an analytic function in the complex plane and it is real for any real argument.
It can also be shown that $h'(y)>0$ in its entire range and hence it is also invertible. 

Then, the solution $y_s(x_s)=h^{-1}(g(x_s))$ of (\ref{continuity2}), which must be monotonically increasing as well (as $g'(x)>0$), can be obtained numerically.
In this way, we can express $\mu'$ in terms of $y_s(x_s)$
\begin{equation}
\label{muprima}
\mu'= \frac{a_s^4}{\alpha}\left(2^{16/3} \alpha^2 P^{4/3}(a_s) y_s^{2/3}(x_s)-\beta^2\right).
\end{equation}
It can be shown from (\ref{continuity2}) that $dy_s/dx_s\geq 1$, 
which implies $y_s(x_s)>x_s$ in its entire range and hence $\mu'>\mu$.

The next step to prove the existence of these solutions is to calculate the difference in the accelerations of the shells immediately after an infinitesimal separation that generates the structure illustrated by figure \ref{vacuumsplitting}. If the accelerations at that point are such that this separation grows with time, which in this case would mean that the acceleration of the central brane is greater than that of the vacuum thin shell, as $r$ decreases away from the brane, then the construction is possible, otherwise it would be forbidden. The equation of motion of the brane (\ref{branemotion}) after the separation moment can be written as
\begin{equation}
\label{braneafter}
H_{b'}^2 = - V_{\mu',1}(a) = \frac{1}{8\alpha}\left[2^{8/3}\alpha P(a)^{2/3}h(y(a)) -2-\frac{8k\alpha}{a^2} \right],
\end{equation}
and from here the acceleration of the shell can be obtained as
\begin{eqnarray}
&&\ddot{a}_{b'}(a)=\frac{1}{2}\frac{d}{da} (a^2H_{b'}^2) \nonumber \\
\label{accelbrane} &&=-\frac{a}{4\alpha}\left[1+2^{5/3}\alpha P^{2/3}(a)\left(1+\frac{a}{3}\frac{P'(a)}{P(a)}\right)\left.\left(3y\frac{dh}{dy}-h\right)\right|_{y=y(a)}-\frac{3\beta^2}{2^{11/3}\alpha P^{2/3}(a)}\left.\left(y^{1/3}\frac{dh}{dy}\right)\right|_{y=y(a)} \right]. \nonumber \\
&&
\end{eqnarray}
On the other hand, the acceleration of any of the vacuum shells can be obtained from (\ref{accelvac}), which in this context can be written as
\begin{equation}
\ddot{a}_v(a)=-\frac{a}{4\alpha}\left[1-\frac{\beta^2}{2^{8/3}\alpha P^{2/3}(a)(y^{1/3}(a)-x^{1/3}(a))}\right].
\end{equation}
Although these accelerations are calculated using different time coordinates, it can be shown that $\ddot{a}_v-\ddot{a}_{b'}$ is proportional to the relative acceleration calculated with a single time coordinate defined in the stringy region. Let us consider a $\tau$ coordinate defined within a shell that is a boundary of a region with metric $ds^2=-f(r)dt^2+f^{-1}(r)dr^2+r^2d\Sigma^2_k$, then we can write
\begin{equation}
\left(\frac{d\tau}{dt}\right)^2=\frac{f(a)}{f(a)+\dot{a}^2},
\end{equation} 
where $a(t)$ is the scale factor of the shell at time $t$, and the dot denotes derivative with respect to $\tau$. In this way, the acceleration with respect to $t$ can be written as
\begin{equation}
\label{att}
\frac{d^2a}{dt^2} = \frac{1}{f(a)}\left(\frac{d\tau}{dt}\right)^4\ddot{a} + \frac{f'(a)}{2f(a)}\left(f(a)-\left(\frac{d\tau}{dt}\right)^2\right)\left(2f(a)-\left(\frac{d\tau}{dt}\right)^2\right).
\end{equation} 
Then, if we apply (\ref{att}) to the separation moment, as $\dot{a}$ is the same for both shells at that time, choosing the standard coordinate $t$ corresponding to the stringy region, which it makes sense as long as $f(a_s)>0$, we get
\begin{equation}
\label{propor}
\left.\left(\frac{d^2a_v}{dt^2}-\frac{d^2a_{b'}}{dt^2}\right)\right|_ {a=a_s}=\frac{f^3(a_s)}{(f(a_s)+\dot{a_s}^2)^2}\left.(\ddot{a}_v-\ddot{a}_{b'})\right|_{a=a_s},
\end{equation}
where $f(a_s)=k+(a_s^2/4\alpha)(1+\sqrt{\beta^2+\alpha\mu'/a_s^4}).$\footnote{If it happens that $f(a_s)<0$, which is only possible in the case $k=-1$, then the standard $t$ coordinate would not be timelike, and another time coordinate for the stringy region should be chosen, for example $r$. The proportionality illustrated by (\ref{propor}) would still hold but with a different factor.}

In this way, the difference between the acceleration of the vacuum shell and that of the brane at the moment of separation is proportional to 
\begin{eqnarray}
\ddot{a}_{v}(a_s)-\ddot{a}_{b'}(a_s)=&&-\frac{a_s}{4\alpha}\left[2^{5/3}\alpha P^{2/3}(a_s)\left(1+\frac{a_s}{3}\frac{P'(a_s)}{P(a_s)}\right)\left(h(y_s)-3y_s\frac{dh}{dy}(y_s)\right) + ... \right. \nonumber \\ && \left. ... + \frac{\beta^2}{2^{8/3}\alpha P^{2/3}(a_s)}\left(\frac{3y_s^{1/3}}{2}\frac{dh}{dy}(y_s)-\frac{1}{y_s^{1/3}-x_s^{1/3}}\right)\right].
\end{eqnarray}
Now, using the continuity equation (\ref{continuity}) we can replace 
\begin{equation}
1+\frac{a}{3}\frac{P'(a)}{P(a)} = \frac{\sigma-p(a)}{\sigma +\rho(a)}
\end{equation}
and the condition of existence for this class of solutions ($\ddot{a}_{v}(a_s)-\ddot{a}_{b'}(a_s)<0$) can be written as
\begin{equation}
\label{criterion}
x^{-2/3}_{\infty}\left(1-\frac{p(a_s)}{\sigma}\right)\left(1+\frac{\rho(a_s)}{\sigma}\right)^{1/3}>y_s^{-2/3}\xi(x_s),
\end{equation}
where \footnote{This would not be the limiting value of $x(a)$ if the matter-energy content of the brane is a ``cosmological constant fluid''. In that case $\rho(a)=-p(a)=\lambda$, and $x(a)$ would tend to $\beta^3/(\kappa^4\alpha(\sigma+\lambda)^2)$.}
\begin{equation}
x_{\infty}=\lim_{a \to \infty} x(a) = \frac{\beta^3}{\kappa^4\alpha\sigma^2}  \;\;\; and \;\;\;  \xi(x_s)=\frac{\frac{2y_s^{2/3}}{y_s^{1/3}-x_s^{1/3}}-3y_s \frac{dh}{dy}(y_s)}{h(y_s)-3y_s \frac{dh}{dy}(y_s)}.
\end{equation}
There is a straightforward necessary condition that one can see from (\ref{criterion}): $\sigma>p(a_s)$, which implies that, in a radiation dominated universe, there is a maximum redshift at which this construction can take place, as we shall see.

The only remaining ingredient to prove the plausibility of figure \ref{vacuumsplitting} is to show that the condition (\ref{criterion}) is satisfied in a certain range $a_{min}<a_s<a_{max}$ for certain parameters and equation of state for the matter-energy content of the brane. 
The general strategy, which is not exhaustive, that we apply in order to find specific examples stems from an analysis of $\xi(x_s)$ and $x(a)$. 
We first must determine a positive lower bound for the factor $(1-p(a_s)/\sigma)(1+\rho(a_s)/\sigma)^{1/3}$, call it $b$, which will depend on the matter-energy content and, in general, it will restrict the domain of $a_s$. Then, although the function $\xi(x_s)$ is positive and monotonically increasing, 
the function $y_s^{2/3}(x_s)/\xi(x_s)$ is also monotonically increasing. In this way we demand the function $x(a)$ to acquire values sufficiently greater than $x_{\infty}$ such that we can choose $x_s$ large enough to satisfy $y^{2/3}_s(x_s)/\xi(x_s) > (1/b) x^{2/3}_{\infty}$. This might be done by choosing carefully the parameters within the definition of $x(a)$, which can be written as
\begin{equation}
\label{xinfty}
x(a)=x_{\infty}\frac{\left(1+\frac{a^4_{\mu}}{a^4}\right)^{3/2}}{\left(1+\frac{\rho}{\sigma}\right)^2},
\end{equation}
where $a^4_{\mu}=(\alpha\mu)/\beta^2$. We will return to this discussion when we apply it to the case of a radiation dominated universe. 

On the other hand, by means of (\ref{xinfty}), it can be shown as follows that there can not be solutions if $\mu\leq0$. In that case we would have $a^4_{\mu}\leq0$, which implies $x(a)\leq x_{\infty}/(1+\zeta(a))^2$ (the equality would hold for $\mu=0)$, where $\zeta(a)=\rho(a)/\sigma>0$. At the same time, the dominant energy condition would imply $-\zeta(a)\leq p(a)/\sigma \leq \zeta(a)$. Then, the left hand side of (\ref{criterion}) satisfies the inequalities
\begin{equation}
x^{-2/3}_{\infty}\left(1-\frac{p(a_s)}{\sigma}\right)(1+\zeta(a_s))^{1/3} \leq x_s^{-2/3}\left(1-\frac{p(a_s)}{\sigma}\right)(1+\zeta(a_s))^{-1} \leq x_s^{-2/3}.
\end{equation}
Further, it can be shown that the function $(y_s(x_s)/x_s)^{2/3}-\xi(x_s)$ is negative in its entire range, so $x_s^{-2/3}<y_s^{-2/3}\xi(x_s)$, which in turns implies that {\bf there can not be a solution satisfying (\ref{criterion}) if $\mu\leq 0$}. In particular, this construction is not possible if the initial bulk spacetime is $dS_5$ or $AdS_5$, and from now on we will assume $\mu>0$.

As we shall see, examples can be found, and this kind of construction exists. Nevertheless, as a general expression of (\ref{criterion}) in terms of $a_s$ is complicated, it is useful, in order to gain insight, an analysis of the asymptotic forms of it.

\subsection{Large $a_s$ limit}

For $a_s$ large enough, $x_s\approx x_{\infty}$ and the condition (\ref{criterion}) takes the form\footnote{Again, in the case of a cosmological constant fluid $x_s\approx x_{\infty} \sigma^2/(\sigma+\lambda)^2$. Anyway, the analysis of this subsection would also hold as the effect of considering this case in the criterion (\ref{criterion}) would be the mere replacement of $\sigma$ by $\sigma+\lambda$ while making $\rho(a)=p(a)=0$.}
\begin{equation}
\left(\frac{y_s(x_{\infty})}{x_{\infty}}\right)^{2/3}>\xi(x_{\infty}),
\end{equation}
which is an inequality involving $x_{\infty}$ only. As mentioned, $(y_s(x_s)/x_s)^{2/3}<\xi(x_s)$, so {\bf this construction is not possible for large $a_s$}. In other words, if $\rho(a_s)<<\sigma$, $|p(a_s)|<<\sigma$ and $a_s>>a_{\mu}$, then this construction can not be made. As discussed in appendix \ref{cosmology}, this fact implies (provided our Friedmann equation is set to asymptotically coincide with the $\Lambda$-CDM model) the impossibility of having this kind of splitting at a redshift corresponding to the ``standard model regime''.

\subsection{Small $a_s$ limit}
As discussed, the criterion (\ref{criterion}) can only be satisfied if $p(a_s)<\sigma$. So, for any matter-energy content such that $p'(a)<0$ this solution might not be possible for an arbitrarily small $a_s$. We assume that the function $\omega(a)=p(a)/\rho(a)$ has a limiting value $\omega_0=\lim_{a\to0}\omega(a)$, which implies that for a sufficiently small $a$: $\rho(a)\approx Ca^{-3(1+\omega_0)}$ and $p(a)\approx \omega_0 \rho(a)$. In these terms, $p(a_s)<\sigma$ can only be satisfied for small $a_s$ if $\omega_0\leq0$. This {\bf precludes any linear barotropic fluid with positive pressure}, in particular it precludes a radiation-dominated universe ($\omega_0=1/3$). In this way, the function $x(a)$ would have the following limiting behaviour 
\begin{equation}
x(a) \approx \frac{(\alpha\mu)^{3/2}a^{6\omega_0}}{\kappa^4\alpha C^2},
\end{equation} 
where it can be seen that $x(a)$ tends to a positive constant for $\omega_0=0$ (a matter-dominated universe), and to $+\infty$ for $\omega_0<0$.\footnote{This approximation remains true if $\omega_0=-1$, but we should replace $C$ with $C+\sigma$ in the above expression.} 

On the other hand, if we Taylor expand both sides of (\ref{continuity2}), taking into account (\ref{hextended}), as functions of $u=x_s^{-1/3}$ and $v=y_s^{-1/3}$ respectively around $u=v=0$, it can be seen that for sufficiently large $x_s$ we must have $y_s^{2/3}\approx 4x_s^{2/3}$. Then, for $\omega_0<0$, (\ref{criterion}) can be written as
\begin{equation}
x_{\infty}^{-2/3}\frac{C^{4/3}}{\sigma^{4/3}}|\omega_0| a_s^{-4(1+\omega_0)}> \frac{\xi(0)\kappa^{8/3}\alpha^{2/3}C^{4/3}a_s^{-4\omega_0}}{4\alpha\mu} \Leftrightarrow  a_s^4 < \frac{4\mu|\omega_0|\alpha}{\xi(0)\beta^2}
\end{equation}
which {\bf always holds for sufficiently small $a_s$}.  

In the remaining case, which includes a matter-dominated universe ($\omega_0=0$), $x(a)$ acquires the limiting value $x_0=(\alpha\mu)^{3/2}/(\kappa^4\alpha C^2)$. Then, condition (\ref{criterion}) can be written as
\begin{equation}
x_{\infty}^{-2/3}\frac{C^{1/3}}{\sigma^{1/3} a_s} > y_s^{-2/3}(x_0) \xi(x_0) \Leftrightarrow a_s < \frac{y_s^{2/3}(x_0)C^{1/3}}{x_{\infty}^{2/3}\sigma^{1/3}\xi(x_0)},
\end{equation}
which {\bf always holds for sufficiently small $a_s$}. We then found several scenarios in which the construction is possible: linear barotropic fluids of non-positive pressure, which in particular includes the dust brane (a matter dominated universe), in the limit of small $a_s$. However, in the context of brane-world cosmology none of these situations can be attained if standard cosmology is to be recovered, as explained in appendix \ref{cosmology}. 

\subsection{Radiation dominated universe} 
\label{photongas}

Because of the fact that in the ``standard model regime'' this construction is not possible, one should check whether it can be done for the early universe. We then consider the special case in which the matter-energy content of the brane is a photon gas ($p=1/3\rho$). Looking at the inequality (\ref{criterion}) we notice that the factor $\psi(\zeta)=(1-\zeta/3)(1+\zeta)^{1/3}$ is a monotonically decreasing function of $\zeta=\rho/\sigma$ whose maximum value is $\psi(0)=1$. Then, in order to satisfy (\ref{criterion}) the following inequality must hold
\begin{equation}
\frac{y_s^{2/3}(x_s)}{\xi(x_s)} > x_{\infty}^{2/3}. 
\end{equation}
Inverting the left hand side, this inequality implies that for a given $x_{\infty}$ there is a minimum value for $x_s$, which we may write $x_{s,min}(x_{\infty})>x_{\infty}$ (where $x_{s,min}(x_{\infty})$ is a monotonically increasing function of $x_{\infty}$ that satisfies $x_{s,min}(0)=0$). 
In this way, for a given $x_{\infty}$ and $x_{s}>x_{s,min}(x_{\infty})$, condition (\ref{criterion}) can be written as $\psi(\zeta)>(x_{\infty}/y_s(x_s))^{2/3}\xi(x_s)$, which sets a maximum for $\zeta$ and, consequently, a minimum for $a_s$ (which we might call $a_{s,min}(x_{\infty},x_s)$).

On the other hand, expression (\ref{xinfty}) for this case can be written as 
\begin{equation}
\label{xinftyrad}
x(a)=x_{\infty}\frac{\left(1+\frac{a^4_{\mu}}{a^4}\right)^{3/2}}{\left(1+\frac{a^4_{\sigma}}{a^4}\right)^2},
\end{equation}
where $a_{\sigma}$ is defined through $\rho(a_{\sigma})=\sigma$. If $a^4_{\mu}>(4/3)a^4_{\sigma}$, defining $r=a^4_{\mu}/a^4_{\sigma}$, this expression has a maximum at $a^4_m=a^4_{\sigma}r/(3r-4)$, where it takes the value
\begin{equation}
x_{max}(r,x_{\infty})=x(a_m)=x_{\infty}\frac{3^{3/2}r^2}{16(r-1)^{1/2}}.
\end{equation}
If $r\leq4/3$, then $x(a)$ is a monotonically increasing function that tends asymptotically to $x_{\infty}$. We then can rule out this case: inequality (\ref{criterion}) can not be satisfied if $r\leq4/3$. We stress the fact that the bound $x_{max}(r,x_{\infty})$ is independent from the condition (\ref{criterion}). 

We then must compare the two bounds on $(x_s/x_{\infty})$: $(x_{max}(r,x_{\infty})/x_{\infty})$, which is a monotonically increasing function of $r$ alone; and $(x_{s,min}(x_{\infty})/x_{\infty})$, which is a function of $x_{\infty}$ bounded from below (around $4.35$). 
In this way, in order for these bounds to be compatible with each other, there is a lower bound $r_{min}(x_{\infty})$, and, in particular, {\bf there can only be a solution if $r>5.26$}, which is equivalent to $a_{\mu}>1.51\; a_{\sigma}$. We then found a necessary condition for (\ref{criterion}): 
$r>r_{min}(x_{\infty})>5.26$. If satisfied, then the bounds on $x_s$ define a range for $a_s$ that must contain the range in which (\ref{criterion}) holds, provided it exists. 


Finally, as mentioned, for a given pair $(x_{\infty},x_s)$, (\ref{criterion}) is equivalent to $a_s>a_{s,min}(x_{\infty},x_s)$. If we set $x_{\infty}$ alone, $a_{s,min}(x_{\infty},x_s)$ is a monotonically decreasing function of $x_s$ for $x_s>x_{s,min}(x_{\infty})$, that tends to $+\infty$ in the lower limit and to $3^{-1/4}a_{\sigma}$ when $x_s\to\infty$. Then, there is a range $a_{min}<a_s<a_{max}$ in which (\ref{criterion}) is satisfied if and only if the curves $a=a_{s,min}(x_{\infty},x)$ and $x=x(a)$ intersect in the plane $(a,x)$.
The possibility of this intersection is not self-evident, but it can be proven as follows by giving a concrete example. This will also be useful to illustrate the definitions we introduced in this subsection.

\subsubsection*{Examples with $x_{\infty}=1$}

\begin{figure}[h]
\centerline{\includegraphics[width=.5\textwidth]{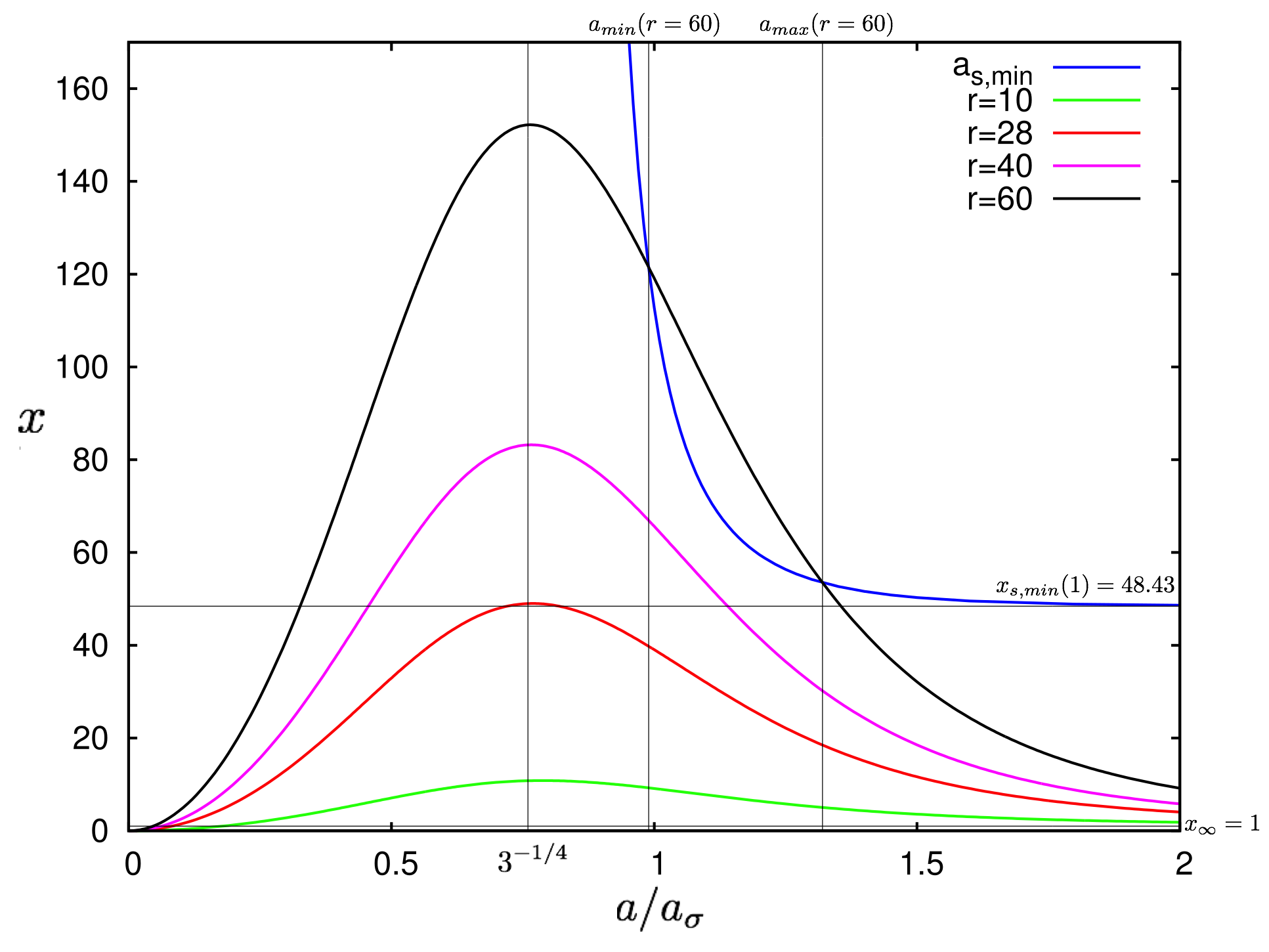}}
\caption{Plot of the function $x(a)$, with $x_{\infty}=1$, for different values of $r$, and of the function implicitly defined by $a=a_{s,min}(1,x)$. The horizontal axis scale is set in terms of $a/a_{\sigma}$. For $r>27.28$ the maximum value of $x(a)$ is greater than $x_{s,min}(1)=48.43$. The curves for $r=10$, $r=28$, $r=40$ and $r=60$ are plotted. Among these curves only the one corresponding to $r=60$ intersects $a_{s,min}(1,x)$, so in that case there is a range for $a_s$ in which (\ref{criterion}) is satisfied, whose boundary values are denoted as $a_{min}(r=60)$ and $a_{max}(r=60)$.}
\label{radiationexample}
\end{figure}

If $x_{\infty}=1$ then $x_{s,min}(x_{\infty}=1)=48.43$. Also, $r_{min}(x_{\infty}=1)=27.78$, so we must choose $r$ greater than that in order to found a solution of (\ref{criterion}). We then plot in the $(a,x)$ plane the function $x(a)$ for different values of $r$ and the function $a=a_{s,min}(x_{\infty}=1,x)$, as illustrated in figure \ref{radiationexample}.

Graphically, it can be seen that an intersection takes place for $r\geq49$, so {\bf there are solutions for a radiation dominated universe}. We illustrate the case $r=60$, in which the intersection points are at $a_{min}(r=60)=0.99 a_{\sigma}$ and $a_{max}(r=60)=1.32 a_{\sigma}$. In this way, if the parameters are such that $x_{\infty}=1$ and $r=60$, which is possible since they are independent ($\mu$ is present only in $r$), then criterion (\ref{criterion}) is satisfied in the range $0.99 a_{\sigma}<a_s<1.32 a_{\sigma}$ and a splitting solution, as illustrated in figure \ref{vacuumsplitting}, can be constructed.

Furthermore, in Section \ref{example} we will provide another example in which this construction holds and such that it tends to the $\Lambda$-CDM universe in the large $a$ limit.




\section{Final outcome of the splitting}
\label{outcome}
Provided that condition (\ref{criterion}) is satisfied for a given set of parameters and some $a_s$, it remains to determine what would be the final outcome of the spacetime after the splitting. The evolution of the central brane would be determined by (\ref{braneafter}) where $\mu'$ is given by equation (\ref{muprima}). Although the evolution equation would be different than before the splitting, it seems difficult for the potential to acquire a point of return, provided the original potential (\ref{branemotion}), with $\xi=-1$, did not exhibit such a property, as expected from brane-world cosmology applications. Nevertheless, the general analysis of the motion allowed by (\ref{branemotion}) is outside of the scope of this paper. On the other hand, the motion of the vacuum thin shells is much easier to describe qualitatively. In principle, there are two different possibilities for the final outcome of the spacetime depending on the fate of the vacuum thin shells: they can either expand indefinitely or collapse\footnote{The possibility of a static vacuum thin shell is precluded because we are always considering $\dot{a}_b(a_s)=\dot{a}_{vac}(a_s)\neq0$.}. As described, the motion of these shells would be determined by the potential (\ref{vaceq}), where $(\xi_+=1,\xi_-=-1,\mu_+=\mu',\mu_-=\mu)$, so the possibility of an indefinite expansion depends on the existence of a point of return and on the relative position of $a_s$ with respect to this point. 

\begin{figure}
\centerline{\includegraphics[width=.25\textwidth]{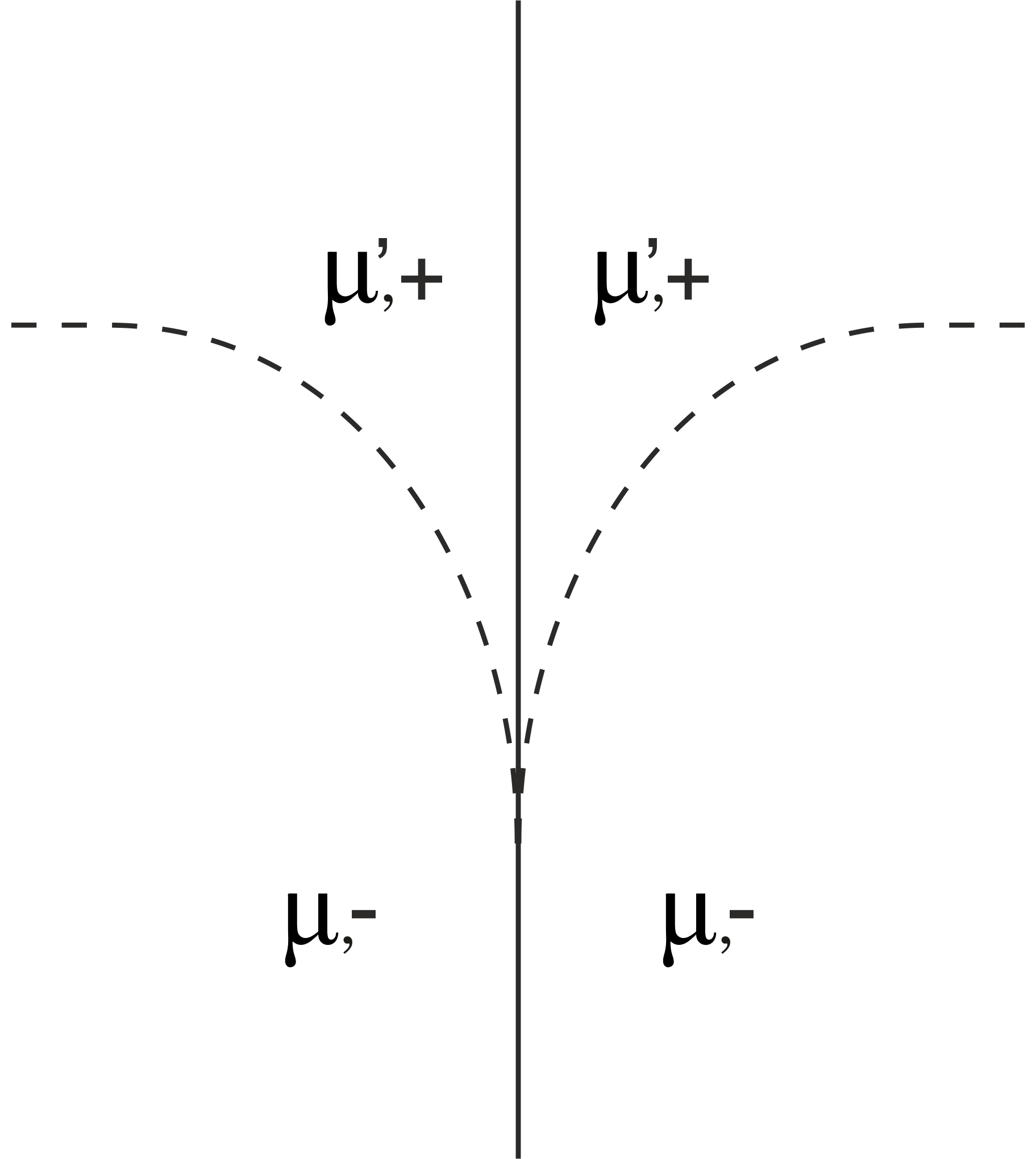}}
\caption{Figure illustrating the possibility of the collapse of the vacuum thin shells. The final outcome would be a bulk spacetime of the stringy type.}
\label{norecoil}
\end{figure}

The possibility of having extremal points for the effective potential (\ref{vaceq}) adapted to this situation can be easily addressed by means of (\ref{accelvac})\footnote{See \cite{garraffo} for a detailed analysis of the potential in the case of a ``false vacuum bubble''.}. From this equation one can see that there must be an extremal point at $a_e$, which is a solution of the expression
\begin{equation}
\label{ae}
A_{\mu'}^{1/2}(a_e)-A_{\mu}^{1/2}(a_e)=\beta^2.
\end{equation}   
The left hand side of this equation is monotonically decreasing with $a_e$ and its image, for positive $a_e$, is $\mathbb{R}^+$, so this equation must have one and only one root. In this way, the effective potential for the motion of the vacuum shell has only one extremal point, and it can be shown that it is a maximum, so the possibility of having points of return is determined by $V_{vac}(a_e)>0$, which is equivalent to the following expression
\begin{equation}
\label{outcomecriterium}
(3-\beta^2-A^{1/2}_{\mu}(a_e))\beta^2-A_{\mu}(a_e)>0.
\end{equation} 
If this inequality holds and $a_s<a_e$, an initially expanding shell (which is always the case in this context) would rebound at some point of the evolution and collapse afterwards. In that case, {\bf the final outcome of the splitting would be a stringy bulk} with mass parameter $\mu'$, as illustrated by figure \ref{norecoil}.  

\begin{figure}
\centerline{\includegraphics[width=.15\textwidth]{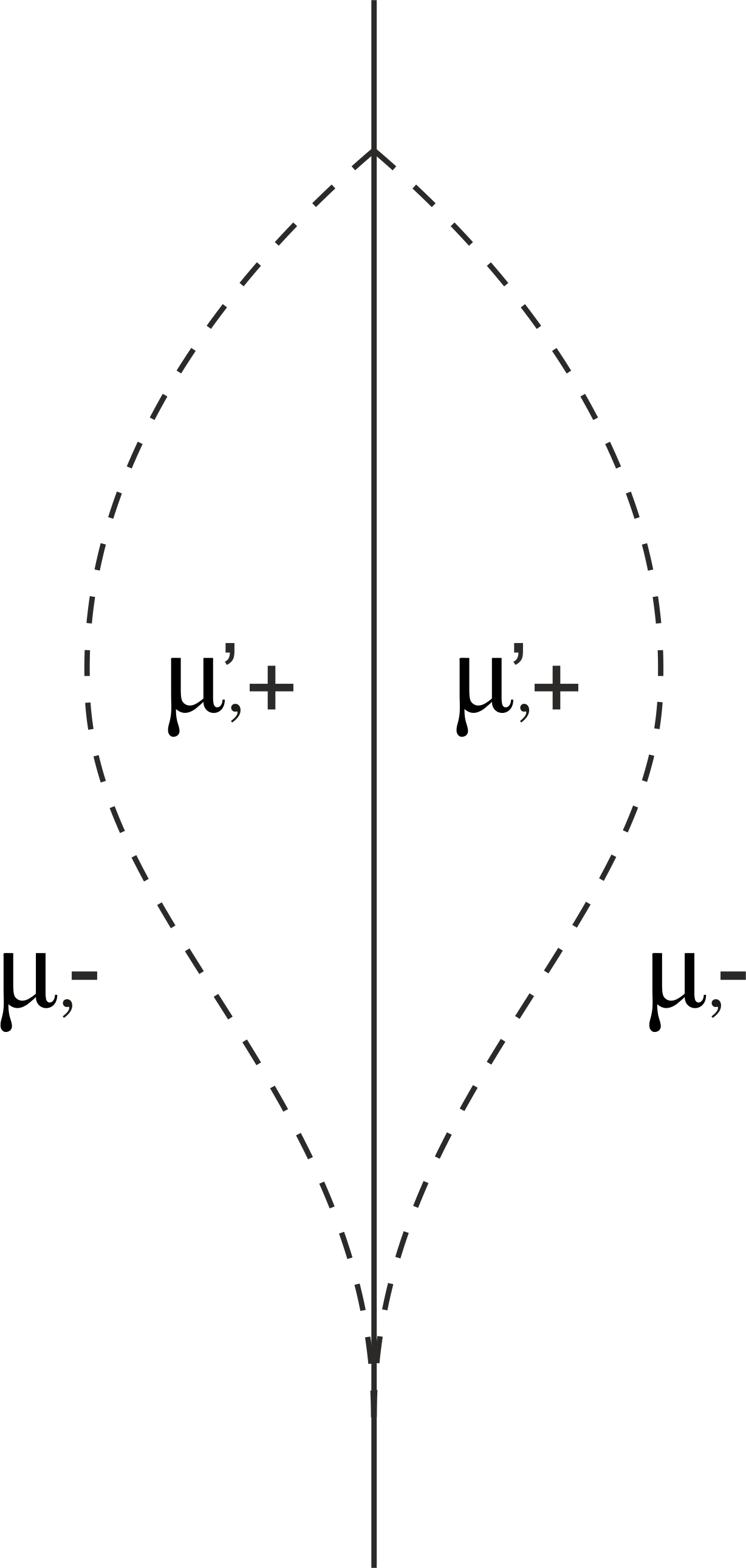}}
\caption{Figure illustrating the possibility of a recoil after the splitting. The final outcome would be a bulk spacetime identical to the original one.}
\label{recoil}
\end{figure}

On the other hand, if $V_{vac}(a_e)\leq0$, there would not be any point of return, so the vacuum shell would expand indefinitely according to (\ref{vaceq}). Anyway, an indefinite expansion, as such, is surely not possible as the shells would eventually recoil. One can notice this by considering the large $a$ limits of (\ref{braneafter}) and (\ref{vaceq}): $\dot{a}_{b'}$ grows like $a^2$ while $\dot{a}_{v}$ grows like $a^6$. In this way, in a scenario where both shells are supposed to indefinitely expand according to their effective potentials, the shells will end up colliding again, as illustrated by figure \ref{recoil}, and a brane collision analysis, like the ones performed in \cite{branecollision}, will play a role in determining the evolution beyond this point.    

The criterion we adopted to determine the splitting parameters cannot be applied to resolve the outcome of the collision in this setting: continuity of $\dot{a}_{b'}$ for the central brane would imply continuity of $\dot{a}_{v}$ as well, which is precluded by the $Z_2$ symmetry. We then must resort to another criterion, which can not be the continuity of the velocity (the tangent vectors of comoving observers within the shells) or the normal vectors, as both coincide and result in the continuity of $\dot{a}_{v}$. We argue that the most reasonable outcome is a recombination of the shells, as illustrated by figure \ref{recoil}, which results in the same bulk spacetime (with the same parameters) as initially. Although this would imply a discontinuity of $H_{b'}$ and of the normal vector of the brane, a further rebound is hard to justify as it would require the introduction of an extra parameter: the initial ``rebound'' velocity of the vacuum shells (or, equivalently, the mass parameter of the stringy spacetime between the brane and the rebounded vacuum shell).      



\begin{figure}
\centerline{\includegraphics[width=.35\textwidth]{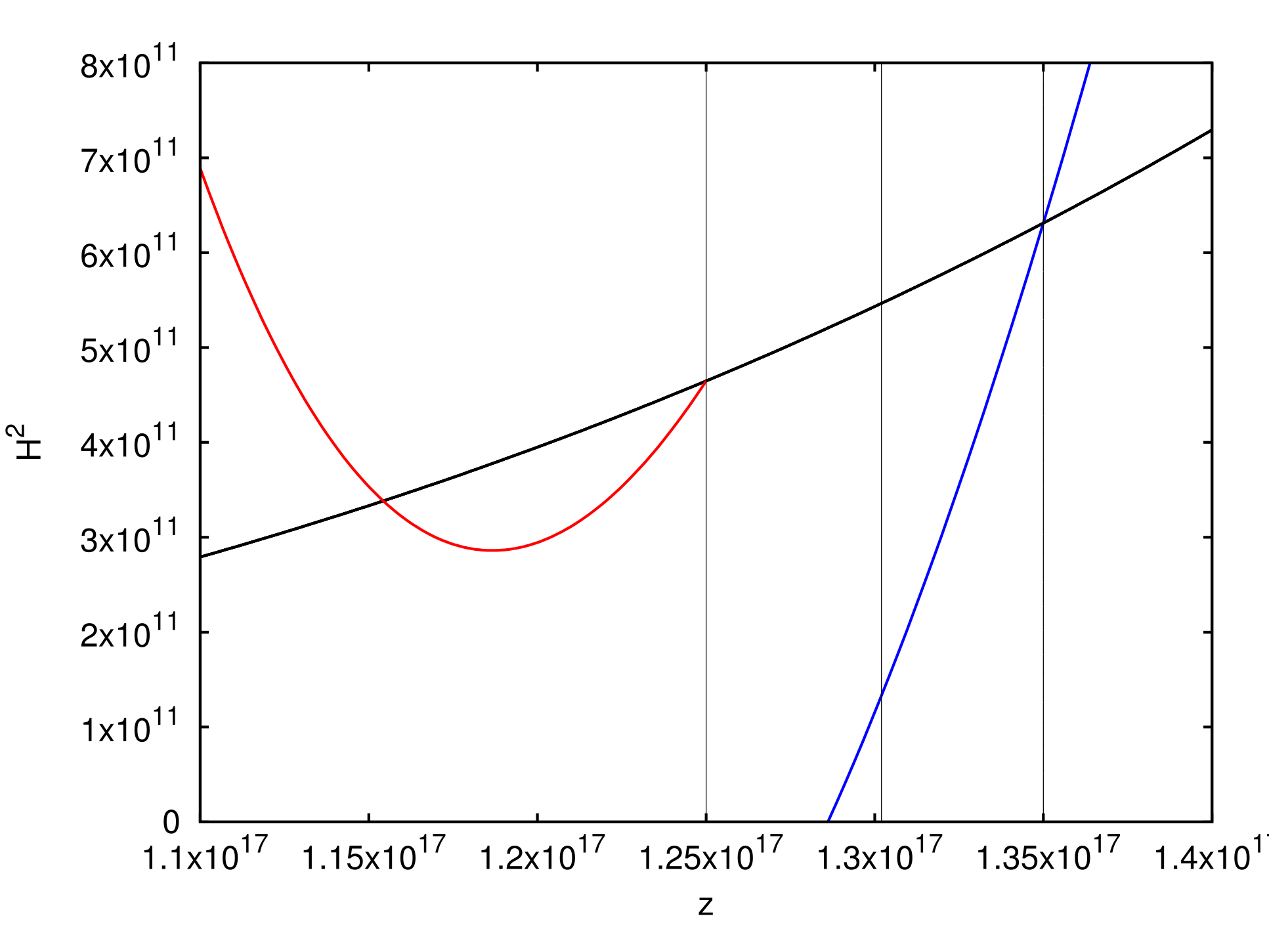}}
\caption{$H^2$ for the brane of the example of Section \ref{example} and for the resulting vacuum shells as functions of $z=a_0/a$ when the splitting takes place at $z_s=1.25\;10^{17}$ and at $z_s=1.35\;10^{17}$. The black line actually represents three different functions: $H^2$ for the original brane and for the resulting ones after the two proposed splittings, but they all practically coincide at this scale. The blue line represents the vacuum shell originating at a splitting at $z_s=1.35\;10^{17}$. It has a point of return, so the shell would contract after it reaches this point and subsequently collapse leaving a stringy bulk. The red line represents the vacuum shell from a splitting at $z_s=1.25\;10^{17}$. It does not have a point of return, so the shell will continue expanding until it recoils with the central brane. The limiting value $z_c=1.302\;10^{17}$ is also marked: if $z_s<z_c$ there would be recoil, while if $z_s>z_c$ the vacuum shell would collapse.}
\label{vacuumpotentials} 
\end{figure}

For a given setting in which (\ref{criterion}) can be satisfied, both final outcomes might be possible for different values of $a_s$. In all the examples we considered (some of them are not included in this paper) this is indeed the case: there is a limiting value for $a_s$, call it $a_c$, which is included in the range satisfying (\ref{criterion}), such that one outcome takes place if $a_{min}<a_s<a_c$ while the other happens if $a_c<a_s<a_{max}$. We illustrate this situation in figure \ref{vacuumpotentials} with the parameters of the example we develop in the next Section.

\section{A cosmological example}
\label{example}


Let us consider a specific example such that the equation of motion of the brane tends to the standard Friedmann equations with $k=0$. In appendix \ref{cosmology} it is explained that provided this asymptotic limit holds, then two of the parameters $(\alpha,\beta,\sigma,\kappa)$ can be written as functions of the other two. We choose $(\alpha,\beta)$ as the independent parameters (besides $\mu$), which implies that $(\kappa,\sigma)$ can be calculated from (\ref{parameters}). Anyway, $(\alpha,\beta,\mu)$ can not be arbitrary, they must satisfy the restrictions (\ref{restriction1}) and (\ref{restriction2}) in order to recover standard cosmology since at least nucleosynthesis and to satisfy observational bounds on dark radiation respectively. 
The scale factor of the universe {\it per se} is not observable, so in order to construct this example we are going to express the dynamics in terms of cosmological redshift. In this way, we should consider $(\mu/a_0^4)$ as the mass parameter to be set, where $a_0$ represents the present scale factor, then write $A_{\mu}(z)=\beta^2+(\mu/a_0^4)(1+z)^4$ and $\rho(z)$ as described in (\ref{rho}), and finally replace $A_{\mu}$ and $P$ in (\ref{branemotion}) and (\ref{vaceq}) with these expressions.  
On the other hand, we must consider the necessary condition to satisfy (\ref{criterion}) that we derived in the Subsection \ref{photongas}.
We then define
\begin{equation}
z_{\sigma}= \frac{a_0}{a_{\sigma}}-1
\;\;\; , \;\;\; z_{\mu}=\frac{a_0}{a_{\mu}}-1=\left(\frac{a_0^4}{\mu}\right)^{1/4}\frac{\beta^{1/2}}{\alpha^{1/4}}-1,
\end{equation}
and choose appropriate values $(\alpha,\beta,z_\mu)$ such that (\ref{restriction1}) and (\ref{restriction2}) hold
and $(z_{\sigma}(\alpha,\beta)+1)>1.51(z_{\mu}+1)$. 

In order to make calculations simpler we first set $z_{\mu}=10^{17}$, so if we are going to impose $z_{\sigma}>1.51\;z_{\mu}=1.51\;10^{17}$, then, from (\ref{rho}) and considering that all species of the standard model are relativistic in this regime, $z_{\sigma}$ can be written as
\begin{equation}
\label{zsigma}
z_{\sigma}^4(\alpha,x_{\infty})=\frac{8\pi\sigma(\alpha,x_{\infty})}{3H_0^2(0.388)\Omega_r}=\frac{\left(\frac{4}{3}\alpha\Lambda_4+1\right)D(x_{\infty})}{6H_0^2(0.388)\Omega_r\alpha g(x_{\infty})},
\end{equation} 
where, as mentioned, $x_{\infty}$ is actually a function of $(\alpha,\beta)$ obtained from (\ref{xinftyab}) 
and $D(x)$ is defined in (\ref{D}).
Now that $z_{\mu}$ is set, we must find a pair $(\alpha,\beta)$ that satisfy all the conditions we mentioned in the above paragraph.
A pair that does the job is $(\alpha,\beta)=(10^{-14}m^2,0.01)$, so the parameters are finally set as follows.
\begin{itemize}
\item $\alpha=10^{-14} m^2$, $\beta=0.01$, $\sigma=1.989\;10^{12} m^{-2}$, $\kappa=3.171\;10^{-3} m^{1/2}$, $z_{\mu}=10^{17}$.
\end{itemize}
Then we get $z_{\sigma}=3.058 \;10^{17}$. For these parameters it turns out that {\bf condition (\ref{criterion}) is satisfied in the range $1.20\;10^{17}<z_s<3.98 \; 10^{17}$}, so the construction illustrated in figure \ref{vacuumsplitting} can be made for any value of $z_s$ within this range.

As explained in Section \ref{outcome}, for a given $z_s$ we can determine the final outcome of the splitting by means of (\ref{ae}) and (\ref{outcomecriterium}). 
We first need to solve (\ref{ae}) in terms of $z_e=(a_0/a_e)-1$,  
and for that one must determine $(\mu'/a_0^4)$, which can be written as a function of $z_s$ as follows
\begin{equation}
\frac{\mu'}{a_0^4}=\frac{1}{\alpha(1+z_s)^4}\left(2^{16/3} \alpha^2 P^{4/3}(z_s) y_s^{2/3}(x_s)-\beta^2\right).
\end{equation}
In this way, for this example one can numerically obtain $z_e(z_s)$ from (\ref{ae}) in the range $1.20\;10^{17}<z_s<3.98 \; 10^{17}$, and then replace it in (\ref{outcomecriterium}). It turns out that $z_e<z_s$ in the entire range, and if $z_s<z_c=1.302 \; 10^{17}$ then the shells will recoil and the final outcome of the splitting is the illustrated in figure \ref{recoil}, while if $z_s>z_c$ then the final outcome is a stringy bulk as in figure \ref{norecoil}. The different possibilities are illustrated in figure \ref{vacuumpotentials}, where $H^2$ for the resulting vacuum shells corresponding to $z_s=1.25\;10^{17}$ and to $z_s=1.35\;10^{17}$ are plotted as functions of $z$.
However, there are reasons to avoid a stringy bulk as a final outcome, as it is well-known that this branch poses instabilities against perturbations \cite{charmousispadilla}, so one may simply preclude this scenario. In any case, a study of the instability of this family of solutions is outside of the scope of the present paper.

We then obtained a concrete example in which the construction can be made, and so that it tends to standard cosmology at low redshift. The redshift at separation $z_s$ can be chosen within a certain range, and both final outcomes are possible depending on this choice. 


\section{Concluding remarks}
\label{conclusions}

In this work we obtain a new class of solutions in Einstein-Gauss-Bonnet gravity, which involves a braneworld in a $Z_2$-symmetric setting from which a pair of vacuum thin shells emanate. The possibility of this construction is non-trivial: it can only be done if the matter-energy content of the braneworld, its scale factor at the splitting point and the parameters of the bulk satisfy (\ref{criterion}). In particular, it is not possible in a regime approximating standard cosmology, or for an arbitrarily small scale factor in a radiation dominated universe. Nevertheless, there are examples that tend to standard cosmology at late times and satisfy (\ref{criterion}) at early, but not arbitrarily early, times.
Of particular interest is the case in which the splitting shells recoil, as illustrated by figure \ref{recoil}. In this case, the bulk spacetime at both sides of the central shell is the same before the splitting than after the recoil, but different for an interval of time in which the bulk is stringy, whose extension depends on the parameters of the construction. During this particular phase of the evolution of the braneworld, the dynamics changes and, in a case developed to emulate the $\Lambda-CDM$ universe at late times, this may affect the termal history of the universe. As mentioned, this mechanism may only play a role in the early universe.
One then may speculate with the consequences of having a sudden change in the acceleration of the rate of expansion, for example in baryogenesis \cite{baryogenesis}, leptogenesis \cite{leptogenesis} or inflation \cite{inflation}, but these are outside of the scope of the present paper, and a matter of future research.

The existence of these solutions is an interesting mathematical fact by itself, because it might represent a drawback against the uniqueness in the initial value problem involving thin shells for Lovelock gravity. Anyway, as illustrated in \cite{ramirez}, this kind of splitting solution also exists in general relativity, so one can argue that this non-uniqueness is more related to the definition of thin shells than to the structure of the EGB field equations. The main difference with respect to the GR splitting solutions is the very existence of vacuum thin shells. For the GR case there must be two different matter-energy fields constituting a single thin shell, and the splitting solution consists on the smooth separation of these constituents. On the other hand, in Lovelock gravity there is no need to separate two different matter-energy fields, one might just consider a vacuum thin shell emanating from a given non-vacuum thin shell, and it turns out that this is possible in a non-trivial way. One possible reason against the naturalness of the constructions made in \cite{ramirez} is the lack of a {\it triggering mechanism} for the splitting, as one may just deem {\it more natural} a single evolving thin shell than the resulting evolution after an infinitesimal separation of the constituent fields. Although this argument is contentious, it is worth noticing that in this case this potential shortcoming is not present, as there is no need to ``arbitrarily'' separate two matter-energy fields to construct the splitting: there is no matter-energy ``leaving'' the original thin shell.   

On the other hand, in the last few years there has been interest in deriving solutions with vacuum thin shells in the context of the thermodynamical instability of vacuum solutions of Lovelock theory. As we mentioned in the case of EGB gravity, a vacuum thin shell can be interpreted as a ``false vacuum bubble'', which is an interface between two different vacua of the theory. There are analogous solutions for higher dimensional Lovelock theory, in which the isotropic vacuum solutions described in (\ref{metric}) are generalised, that also display different branches. Depending on the parameters of the theory, there are up to $K$ branches, where $K$ is the order of the higher order factor in curvature of the field equations \cite{camanhoedelstein1}, and all possible pairs of different branches can possibly be glued with a vacuum thin shell, constituting, in this way, many different types of vacuum bubbles. The static vacuum bubbles can be analysed thermodynamically by Euclidean methods and the {\it transition probability} among the different vacuum solutions can be semiclassically addressed \cite{camanhoedelstein2,hennigarmannmbarek}. For a given set of boundary conditions, the ``true vacuum'' corresponding to them can be singled out by this method. In particular, a ``metastable'' solution may ``thermally'' decay to a bubble configuration and then to a ``true vacuum'' via classical dynamics. This analysis is important in the context of the AdS/CFT correspondence, because it reveals an intricate and previously unknown behaviour of gravity theories that should be replicated in the dual CFTs.

Furthermore, there have also been some studies of this sort involving non-vacuum static solutions \cite{giribetgoyaoliva} with a self-gravitating conformal scalar field, but not, as far as we know, involving non-vacuum thin shells. This latter possibility should be of interest in the quest of exploring the most stable solutions that can be interpreted as final outcomes of the evolution of different types of matter-energy configurations and is a matter of future research. We also remark that, although the solutions considered here are thought as dynamical, the method we developed in this paper is perfectly applicable to a static case outside of the context of braneworld cosmology. We then might interpret the present work as the foundation of a different kind of stability analysis for thin shells in Lovelock gravity, which adds to perturbation analysis and thermodynamical stability analysis. In this way, a generalisation of (\ref{criterion}) for non-$Z_2$-symmetric settings, which would be more algebraically involved, as illustrated in \cite{shtanov}, and a comparison with other types of stability analysis, are also a matter of future research. 

\section{Acknowledgements}
The author acknowledges Ernesto Eiroa for interesting comments and reading the whole manuscript. He also thanks the referees for pointing out a non-trivial mistake and other useful comments. MAR is supported by CONICET. 

\appendix

\section{Existence of the equation of motion for the central brane}
\label{derivation}

Here we basically reproduce the steps taken by Maeda in \cite{maeda}, for the case of a ``GR'' bulk, to demonstrate the existence of an effective potential for the evolution of the brane-world, but considering any of the branches of the bulk solution, as expressed in (\ref{branemotion}). Considering (\ref{rhosq}) we define
\begin{equation}
F(x)=\left(\frac{f(a)}{a^2}+x\right)\left(1+4\alpha\left(\frac{k}{a^2}-\frac{f(a)}{3a^2}+\frac{2}{3}x\right)\right)^2-\frac{\kappa^4}{36}(\rho+\sigma)^2.
\end{equation}
In this way, the Friedmann equations can be written as $H^2=x_s$, where $F(x_s)=0$. We are going to prove that $F(x)$ must have one, and only one, positive real root, so the effective potential always exists and is unique. If we calculate $dF/dx$ one can see that it has two roots
\begin{equation}
x_{\pm}=-\left(\frac{1}{4\alpha}+\frac{k}{a^2}\right)\pm\frac{1}{2}\left(\frac{f(a)}{a^2}-\left(\frac{1}{4\alpha}+\frac{k}{a^2}\right)\right).
\end{equation}    
Furthermore, evaluating $F$ in these two extremal points we have $F(x_+)<0$ and $F(x_-)<0$. As a third degree polynomial in $x$, the principal term defines the asymptotic behaviour of $F(x)$ and it turns out that its coefficient is positive. In this way, $dF/dx$ is positive at both extremes $\pm\infty$, which means that the lesser of the extremal points $x_{\pm}$ must be a local maximum. But, as we commented, $F$ is negative there, so {\bf there will be only one real root at $x_s>x_{\pm}$}. 

Although there is always a first order ordinary differential equation for $a(\tau)$ (\ref{branemotion}) that it is equivalent to (\ref{rhosq}), it is a priori not granted that this equation is equivalent to the $\tau\tau$ component of (\ref{junction}) because of the squaring we performed to get (\ref{rhosq}). There is the a condition that must be satisfied for (\ref{branemotion}) not to be spurious, 
namely
\begin{equation}
sign\left(\left.\frac{\partial r}{\partial \eta}\right|_{\eta=0^+}\right)sign(\rho(a)+\sigma)=-sign\left(a^2+4\alpha\left(k+\frac{2}{3}\dot{a}^2-\frac{1}{3}f(a)\right)\right).
\end{equation}
By means of (\ref{branemotion}), this condition can be written as
\begin{equation}
\label{orientation}
sign\left(\left.\frac{\partial r}{\partial \eta}\right|_{\eta=0^+}\right)=sign(P)sign\left(\xi A^{1/2}_{\mu}-B_{\xi}(P^2,A^{3/2}_{\mu})-\frac{A}{B_{\xi}(P^2,A^{3/2}_{\mu})}\right).
\end{equation} 
Under our assumptions $\rho+\sigma>0$, 
then it is clear that if the bulk is GR then it must also be interior, that is, each bulk region can be defined by an inequality $r<a(t)$. On the other hand, if the bulk is stringy, in order to have $B_{1}(P^2,A^{3/2}_{\mu})$ well-defined, $2^{7/3}\alpha P^{2/3}\geq A_{\mu}^{1/2}$ must hold. But at the same time, by definition, $B_{1}(P^2,A^{3/2}_{\mu})\geq2^{7/3}\alpha P^{2/3}$, so {\bf the bulk must be interior regardless of the value of $\xi$}.
In the same way, one can see that if we allowed $\rho+\sigma<0$ then the bulk should be exterior also regardless of the value of $\xi$. Nevertheless, as explained in appendix \ref{cosmology}, this last possibility forbids the emergence of standard cosmology in the large $a$ limit. 


\section{Large $a$ asymptotics of the effective potential for the central brane}
\label{asymptotics}

The potential in the equation of motion of the shell can then be written as 
\begin{equation}
\label{potentialgr}
V_{\mu,-1}(a)=-\frac{1}{8\alpha}\left[2^{8/3}\alpha P(a)^{2/3}g(x) 
-2-\frac{8k\alpha}{a^2} \right].
\end{equation}
so we need to find the large $a$ limit of $P(a)$ and $x(a)$. The matter-energy degrees of freedom within the braneworld appear in the effective potential through the function $P$. If we impose that the matter-energy content satisfies the dominant energy condition, but it is not (or it does not contain) a cosmological constant fluid, then $\rho\to 0$ when $a\to\infty$ and we can write
\begin{equation}
P^2(a)\approx \frac{\kappa^4}{2^8\alpha^2}\sigma^2+\frac{\kappa^4}{2^7\alpha^2}\sigma\rho(a) \;\;\; , \;\;\;  x(a)\approx x_{\infty}
+\frac{3\beta\mu} {2\kappa^4\sigma^2a^4} - \frac{2\beta^3\rho(a)}{\kappa^4\alpha\sigma^3}.
\end{equation}
Then, linearising the left hand side of (\ref{potentialgr}) as a function of $(P^2,x)$ around the values $((\kappa^4\sigma^2)/(2^8\alpha^2),x_{\infty})$, we obtain the following expression
\begin{eqnarray}
V_{\mu,-1}(a) &\approx& -\frac{1}{8\alpha}((\kappa^4\sigma^2\alpha)^{1/3}g(x_{\infty})-2)-\left(\frac{\kappa^4}{\alpha^2\sigma}\right)^{1/3}\frac{D(x_{\infty})}{12}\rho(a)- \nonumber \\
\label{largealimit}
&&-\frac{(\kappa^4\sigma^2\alpha)^{1/3}(g(x_{\infty})-D(x_{\infty}))\mu}{16\beta^2a^4} +\frac{k}{a^2},
\end{eqnarray}
where
\begin{equation}
\label{D}
D(x)=\frac{1}{(1+2x)^{1/2}}\left(\omega(x)-\frac{x^{2/3}}{\omega(x)}\right).
\end{equation}
As both the functions $D(x)$ and $g(x)-D(x)$ are non-negative for $x>0$, we can see that the equation of motion $H^2=-V_{\mu,-1}(a)$ tends asymptotically to a form similar to the standard Friedmann equations but with an additional term whose effect in the dynamics would be as if there where a radiation density not included in $\rho$ (the so-called {\it dark radiation}).

\section{How to generate our universe}
\label{cosmology}
Looking at (\ref{largealimit}), in order to recover the standard Friedmann equations, in geometric units, the following identifications must be made
\begin{equation}
\label{identifications}
\Lambda_4=\frac{3}{8\alpha}((\kappa^4\sigma^2\alpha)^{1/3}g(x_{\infty})-2) \;\;\; , \;\;\; \frac{\kappa_4^2}{3}=\frac{8\pi}{3}=\left(\frac{\kappa^4}{\alpha^2\sigma}\right)^{1/3}\frac{D(x_{\infty})}{12},
\end{equation}
where we recall $x_{\infty}=\beta^3/(\kappa^4\sigma^2\alpha)$. In this way, the matter-energy content of the brane does not need to include dark energy, as it appears as a consequence of the setting. These identifications justify the assumption $\sigma>0$ and allow us to express two of the parameters $(\alpha,\beta,\kappa,\sigma)$ in terms of the other two. 
If we choose $\alpha$ and $\beta$ 
as the independent ones, we should recall that both parameters are positive and express the other two as functions of them ($\sigma(\alpha,\beta)$ and $\kappa(\alpha,\beta)$). After some manipulation of the relations (\ref{identifications}) we can express $x_{\infty}(\alpha,\beta)$ implicitly by means of the equation
\begin{equation}
\label{xinftyab}
g(x_{\infty})x_{\infty}^{-1/3}=\frac{\frac{8}{3}\alpha\Lambda_4+2}{\beta}.
\end{equation}
From this expression it can be seen that $x_{\infty}$ grows with $\beta$ and decreases with $\alpha$, although the dependence with $\alpha$ is only significant if ${\cal O}(\alpha)>10^{51}$. Also from (\ref{xinftyab}) it is deduced, because of the fact that the left hand side asymptotically approaches $2$ for large $x_{\infty}$, that (\ref{identifications}) can only be satisfied if $\beta<1+(4/3)\alpha\Lambda_4$. Then, we can obtain the other parameters as follows
\begin{equation}
\label{parameters}
\kappa^4(\alpha,x_{\infty})=\frac{\left(\frac{4}{3}\alpha\Lambda_4+1\right)2^{11}\pi^2\alpha}{D(x_{\infty})g(x_{\infty})} \;\;\; , \;\;\; \sigma(\alpha,x_{\infty})=\frac{\left(\frac{4}{3}\alpha\Lambda_4+1\right)D(x_{\infty})}{16\pi\alpha g(x_{\infty})}.
\end{equation}


It can be seen from equation (\ref{xinfty}) that the approximation (\ref{largealimit}) would hold only if $\rho(a)<<\sigma$ and $a>>a_{\mu}$. The termal history of the universe according to standard cosmology
predicts very well different aspects of the observed universe, in particular the primitive abundances \cite{nucleosynthesis}, so we require that this approximation should be valid at least since nucleosynthesis, more specifically since the neutron freeze-out at ${\cal O}(z)=10^{10}$. In this way, in the framework of standard cosmology, whenever this limit does not apply, the matter-energy content of the braneworld is essentially pure radiation, so if we want to describe the dynamics of the early universe then the matter-energy content should be written as \cite{degrees}
\begin{equation}
\label{rho}
\rho(z)=\frac{3H_0^2}{8\pi}\frac{1.84g_{*}(T)}{g^{4/3}_{*s}(T)}\Omega_r z^4 \;\;\; , \;\;\; p(z)=\frac{1}{3}\rho(z),  
\end{equation}
where $g_*(T)$ is the number of relativistic degrees of freedom at a given temperature, $g_{*s}(T)$ is the number of effective degrees of freedom in entropy at the same temperature, and $\Omega_r$ includes the neutrino energy density\footnote{As obtained from $z_{eq}$. Taking the value of this parameter from Planck \cite{planck} we get $\Omega_r=9.16\;10^{-5}$.}. Both $g_*$ and $g_{*s}$ are implicit functions of $z$ as well ($T(z)$ is obtained from $g_{*s}(T)T^3/z^3=cst$) and are independent from the Friedmann equations. According to the standard model of particle physics, for ${\cal O}(T)<10 keV$, we have $1.84g_{*}(T)g^{-4/3}_{*s}(T)=1$; on the other hand, for ${\cal O}(T)>100 GeV$, $1.84g_{*}(T)g^{-4/3}_{*s}(T)=1.84g^{-1/3}_{*}(T)=0.388$. In between, this coefficient decreases with $T$ \cite{degrees}, but with a much slower rate than the growth of $z^4$. In this way, using the cosmological parameters best fit from Planck \cite{planck}, we demand 
\begin{equation}
\label{restriction1}
\sigma(\alpha,\beta)>>\rho(z=10^{10})=4.94 \;10^{-18} m^{-2}  \;\;\; , \;\;\;  z_{\mu}=\left(\frac{\beta^2}{\alpha\mu}\right)^{1/4}a_0>>10^{10},
\end{equation}
where $a_0$ is the present scale factor. 


On the other hand, 
there is one deviation from standard cosmology that is a part of the dominant term in the radiation era: the dark radiation term  
\begin{equation}
H_0^2\Omega_{dr}=\frac{(\kappa^4\sigma^2\alpha)^{1/3}(g(x_{\infty})-D(x_{\infty}))\mu}{16\beta^2a_0^4}. 
\end{equation}
This must be limited as a small fraction of the estimated radiation density parameter $\Omega_r=9.16\;10^{-5}$. 
Then, from (\ref{parameters}) we demand
\begin{equation}
\label{restriction2}
\Omega_{dr}(\alpha,\beta,x_{\infty},z_{\mu})=\frac{\beta}{16\alpha}\frac{(g(x_{\infty})-D(x_{\infty}))x^{-1/3}_{\infty}}{z^4_{\mu}H^2_0}<<10^{-4}.
\end{equation}

We must then choose $(\alpha,\beta,z_{\mu})$ such that (\ref{restriction1}) and (\ref{restriction2}) hold. One can notice that the restrictions are compatible with each other: for $\alpha$ sufficiently small, $x_{\infty}$ is essentially a monotonically decreasing function of $\beta$ only, and, for fixed $\beta$, $\sigma$ can be made arbitrarily large. On the other hand, for $z_{\mu}$ sufficiently large and fixed $\alpha$ and $\beta$, $\Omega_{dr}$ can be made arbitrarily small. In this way, for a given value of $\beta$ we choose a sufficiently small $\alpha$ in order to satisfy (\ref{restriction1}), and then choose a sufficiently large $z_{\mu}>10^{10}$ in order to satisfy (\ref{restriction2}).

\end{document}